\documentclass[12pt]{article}
\usepackage{times}
\usepackage{geometry}
\usepackage[utf8]{inputenc}
\usepackage{enumitem,amssymb}
\usepackage{ragged2e}
\usepackage{graphicx}
\usepackage[numbers,sort&compress]{natbib}
\usepackage{wrapfig}
\usepackage{floatrow}
\usepackage{booktabs}
\usepackage{multirow}
\usepackage{threeparttable} 
\usepackage{parskip}
\usepackage{titlesec}
\titlespacing*{\section}{0pt}{2pt plus 1pt minus 1pt}{1pt}
\titlespacing*{\subsection}{0pt}{2pt plus 1pt minus 1pt}{0.5pt}
\titlespacing*{\subsubsection}{0pt}{1pt plus 1pt minus 1pt}{0.5em}
\titleformat{\subsection}[runin]{\normalfont\bfseries}{}{0em}{}[~\quad]
\titleformat{\subsubsection}[runin]{\normalfont\itshape}{\thesubsubsection}{0em}{}[\quad]

\newlist{thematic}{itemize}{8}
\setlist[thematic]{label=$\square$}
\usepackage{pifont}

\newcommand{\ion}[2]{#1\,{\small\textsc{#2}}}
\usepackage{hyperref}

\begin{document}
\thispagestyle{empty} 
\RaggedRight
\huge
Building a Roadmap for Hubble Science into the 2030s \linebreak

A Decade to Map the Diffuse Universe: FRB--QSO Pairs with HST/COS Spectroscopy \linebreak
 
\normalsize
  
\textbf{Principal Author:}

Name:	Jessica Werk
 \linebreak						
Institution:  University of Washington 
 \linebreak
Email: jwerk@uw.edu

\textbf{Co-authors:} Matthew McQuinn (U Washington); J. Xavier Prochaska 
(UCSC); Sunil Simha (U Chicago); Rongmon Bordoloi (NC State); Liam Connor 
(Harvard); Andrew Fox (STScI); J. Chris Howk (University of Notre Dame); Cameron Hummels (Caltech); Lordrick Kahinga (UCSC); Victoria Kaspi (McGill); Khee-Gan Lee (Kavli IPMU); Nicolas Lehner (University of Notre Dame); Kiyoshi Masui (MIT); Benjamin Oppenheimer (UC Boulder); Vikram Ravi (Caltech); Kate Rubin (San Diego State University); Kirill Tchernyshyov (Rocky Mountain Institute); Yong Zheng (RPI)
\justifying

\textbf{Abstract:}
Jointly analyzing the sightlines of arcsecond-localized fast radio bursts (FRBs) and UV-bright quasars (QSOs) nearby in projection has the potential to provide strong constraints on the phases, mass distributions, and magnetic structure of the diffuse universe. Each probe supplies what the other cannot: FRBs provide integrated electron columns (DM), line-of-sight magnetic field estimates (RM), and scattering constraints ($\tau_{\rm scatt}$) that are independent of gas phase; QSOs provide the redshift- and phase-resolved column densities needed to interpret them. Today, there are only $\sim100$ arcsecond-localized FRBs at $z < 1$, making statistical FRB--QSO pair surveys impossible. By 2035, there will be  $\sim10^{5}$.  Using the most recent FRB localization forecasts and UV-bright QSO catalogs, we estimate that next-generation interferometers will yield thousands of FRB--QSO pairs at angular separations $\theta < 10'$, including $\sim100$ pairs at $\theta < 1'$,  over a common 20,000\,deg$^2$ footprint by 2035. We outline the science enabled by this sample: constraints on CGM ionization fractions and baryon masses;  observational constraints on the role of magnetic fields and turbulence in the CGM and cosmic web; sightline-by-sightline partitioning of the cosmic DM budget; and three-dimensional mapping of the multiphase Milky Way and M31 halos. Together, these measurements directly address the physics of feedback, non-thermal pressure support, and energy balance in the diffuse gas that regulates galaxy growth. HST/COS is the only instrument that can carry out this program, and the 2030s are the only decade in which to do it before Habitable Worlds Observatory (HWO) defines the next era of diffuse universe science.

\pagebreak

\setcounter{page}{1}  
\section*{1.\ Introduction} 
The cycling of matter through the diffuse universe plays a critical role in shaping galaxies and underlies some of the most profound open questions in astrophysics \citep{NationalAcademies2021}. Stars, the luminous tracers of galaxy evolution, account for $\sim$5\% of all baryons \citep[e.g.][]{Behroozi:2010}.  The other $\sim$95\% lie in the multiphase gas of the interstellar medium (ISM), the circumgalactic medium (CGM) and the intergalactic medium (IGM) \citep{McQuinn2014, Tumlinson2017}, where they regulate the fuel supply for star formation and record the thermodynamic history of the cosmic web. The cool and warm ionized phases of this diffuse gas ($T \approx 10^4$ -- $10^{5.5}$ \,K; $n \lesssim 10^{-1}$\,cm$^{-3}$) radiate predominantly in the ultraviolet ($\lambda \approx 100$ -- $3200$\AA), where emission and absorption line diagnostics from a wide range of ionized species constrain gas density, temperature, and metallicity \citep{Bertone2013}.\footnote{While a significant fraction of diffuse baryons may reside in hotter gas ($T \gtrsim 10^6$\,K) whose emission predominantly falls in the X-ray bandpass,  this hot, tenuous phase is still largely inaccessible to current instruments. The compelling science case for a soft X-ray microcalorimeter mission to map O\,{\sc vii} and O\,{\sc viii} emission from the hot CGM and cosmic web \citep[e.g., the \textit{LEM} concept;][]{LEM2022} underscores the gap that UV and radio observatories must fill in its absence.} For this reason, the UV provides the most accessible window to a large fraction of the diffuse gas of the cosmic baryon cycle, making \textit{HST's Cosmic Origins Spectrograph} \citep[COS;][]{Green2012}  the cosmic ecosystems machine of choice for the last 15$+$ years.  

Large programs including COS-Halos, CLASSY, AMIGA, CUBS, ULLYSES, and others \citep[e.g.,][]{Tumlinson2011, 
Werk2013, Berg2022, Lehner2021, Chen20, Roman-Duval2020} have not only established COS as the premier tool in its own right for constraining the physics of galactic atmospheres and winds, but have demonstrated its unique power in combination with data at other wavelengths.  Ground-based integral field spectrographs (e.g., KCWI, MUSE), X-ray observatories (e.g., \textit{Chandra}), and the upcoming UV photon-counting mission \textit{Aspera} can and will work with HST/COS in ways that multiply the scientific return far beyond what any single facility achieves alone. This potential will only grow: ambitious new surveys in the optical (Rubin/LSST), radio (DSA-2000, CHORD), and infrared (\textit{JWST}) will generate enormous demand for UV spectroscopic observations that only \textit{HST}/COS can provide. Losing high-resolution UV access at this moment of convergence could therefore set the astrophysics of the diffuse universe, one of the key drivers of Habitable Worlds Observatory (HWO) science, back by a decade or more.

\section*{2.\ FRBs and QSOs as Complementary Probes}

In this white paper, we outline one science case that will only become 
possible by $\sim$2035, when arcsecond-localized fast radio bursts (FRBs) with confirmed 
spectroscopic redshifts will be sufficiently numerous to enable targeted FRB - Quasar (QSO) pair surveys. The spectra of UV-bright QSOs are threaded with hundreds to thousands of absorption lines arising from intervening gas at a range of redshifts along the line of sight. They offer a sensitive record of the column densities, ionization states, and kinematics of the IGM and CGM encountered between the source and the observer. Yet QSO sightlines have 
fundamental limitations.  In particular, the Ly$\alpha$ forest provides the most complete census of this intervening hydrogen \citep[e.g.][]{Rauch1998}, but traces predominantly neutral gas and is insensitive to the warm ionized phases that dominate the IGM and CGM mass budgets at $z < 1$.  For enriched material, metal-line diagnostics constrain the gas ionization state but require ionization corrections that depend sensitively on uncertain model assumptions \citep{Werk2014}.  FRBs supply this key information missing from QSO absorption-line measurements: the dispersion measure (DM) of an FRB encodes an ionization-model-free, line-of-sight integral constraint on the free-electron column density along the full path from source to observer,
\begin{equation}
    \mathrm{DM} = \int \frac{n_e \, \mathrm{d}s}{1 + z} \, ,
    \label{eq:dm}
\end{equation}

\noindent where $n_e$ is the free-electron number density, 
$\mathrm{d}s$ is the proper path length element, and the factor 
$(1+z)^{-1}$ accounts for cosmological expansion 
\citep{Chatterjee2017, McQuinn2014}. The total measured DM receives contributions from each intervening medium along the sightline: the 
IGM, the CGM and ISM of the FRB host galaxy, the halos of galaxies intersected in projection, and the Milky Way \citep{McQuinn2014, Prochaska2019}. Furthermore, this integral is blind to the gas phase, provided the medium is ionized -- the FRB DM is equally sensitive to cool photoionized gas and the hot, volume-filling phases at 
$T \gtrsim 10^6$\,K that theory predicts dominate the 
baryon budget \citep[e.g.,][]{Cen2006}.  Each FRB also carries a Faraday rotation measure, ${\rm RM} \propto \int 
n_e B_\parallel \, {\rm d}s$, which when divided by the DM yields a density-weighted estimate of the line-of-sight magnetic field through the same plasma, while the pulse broadening timescale $\tau_{\rm scatt}$ constrains 
small-scale electron density fluctuations in intervening gas 
\citep[e.g., turbulence, intermittency, and cloud morphology;][]{McQuinn2014, Jow2026}), probing physical scales 
inaccessible to spectroscopic observations. Both quantities are entirely inaccessible to UV 
absorption-line experiments.


Apportioning the DM and RM of an FRB into contributions from the IGM, intervening galaxy halos, the host galaxy CGM and ISM, and the Milky Way requires assumptions about the gas fraction in each component that currently carry substantial uncertainty \citep[][]{Prochaska2019}. A high-resolution UV spectrum of an adjacent QSO sightline would help break these degeneracies: multi-species absorption lines identify specific structures along the sightline, constrain their hydrogen content (and overdensity), and provide the redshift information needed to partition the integrated FRB DM and RM into contributions from each. FRBs and QSOs therefore supply the key missing observable in each other's experiments, and together they constrain the ionized universe in ways that neither can alone. However, the scarcity of arcsecond-localized FRBs means that such joint analyses are not possible in 2026. The coming decade of dedicated fast-survey interferometers, including DSA-2000 \citep{Hallinan2019}, CHORD \citep{Vanderlinde2019}, and the Square Kilometre Array \citep{Macquart2015}, is expected to deliver hundreds of thousands of localized FRBs with host redshifts, making a statistical FRB--QSO pair survey feasible for the first time by 2035\footnote{The redshift 
distributions of localized FRBs and UV-bright QSOs will 
not generally be matched: for a given angular pair, the 
QSO may lie in front of or behind the FRB host. Both 
geometries are scientifically useful as long as they are 
known from spectroscopic redshifts.}. However, the UV spectroscopic capability to realize the scientific potential of these FRB sightlines depends on the future of \textit{HST}/COS.

\section*{3.\ Forecasting the FRB--QSO Pair Sample by 2035}
To estimate the number of UV-bright QSO--FRB angular pairs, we compute the expected cumulative pair count within projected angular separation $\theta$ as $N(<\theta) = N_{\rm FRB} \,\Sigma_{\rm QSO}\, \pi\theta^2$, where $N_{\rm FRB}$ is the estimated number of arcsecond-localized FRBs at $z < 1$ within the survey footprint, $\Sigma_{\rm QSO}$ is the surface density 
of UV-bright QSOs (per deg$^{-2}$), and $\pi\theta^2$ is the solid angle subtended by a circle of angular radius $\theta$. This expression, by design, assumes a Poisson 
random field, and since the FRB and QSO redshift distributions are largely uncorrelated, large-scale clustering provides negligible boost to the pair counts.  We 
evaluate $N(<\theta)$ over an assumed common survey footprint of  $\sim$20{,}000~deg$^{2}$.

The UV-bright QSO surface density $\Sigma_{\rm QSO}$ is estimated from the 
Milliquas v8 catalog \citep{Flesch2023} cross-matched with GALEX photometry catalogs, and selecting 
QSOs at $z < 1$ with FUV $< 19$~AB.  This flux threshold corresponds roughly to $\mathrm{S/N} \gtrsim 10$ 
per resolution element achieved in $\lesssim$15 orbits per target with COS (G130M $+$ G160M; e.g. COS.sp.2336278). Our cross-match yields a fiducial surface density of $\Sigma_{\rm QSO} \approx 0.46$~deg$^{-2}$, with a plausible range of $0.38$--$0.54$~deg$^{-2}$. For the present-day ($\sim$2026) epoch, we adopt $N_{\rm FRB} \approx 100$ arcsecond-localized FRBs at $z < 1$, consistent with the $\sim$90 reported by \cite{Connor2025} and accounting for ongoing growth; this estimate is conservative. By 2035, next-generation interferometers are expected to deliver $N_{\rm FRB} \sim 10^5$ localized FRBs at $z < 1$ \citep{Hallinan2019, Vanderlinde2019, Sharma2026} , which we adopt as our fiducial 2035 value, with a range of $\sim$48{,}000--135{,}000 spanning conservative to optimistic scenarios. Dedicated deep-drilling fields, such as those planned for the DSA-2000 over regions like COSMOS where areal FRB densities may reach $\sim$200\,deg$^{-2}$, would substantially boost the number of close pairs and enable the highest-impact science cases described below. Additionally, we treat the UV-bright QSO catalog as static for this estimate, although this is almost certainly conservative.  Rubin/LSST is expected to detect of order tens of millions of AGN and quasars across the full survey  footprint \citep{Li2026}, and the resulting increase in known QSOs, some of which will be UV-bright, will only push $\Sigma_{\rm QSO}$ higher and further increase pair counts.

\begin{figure}[h!]
\begin{centering}
\vspace{-0.5cm}
\hspace{-0.8in}
  \floatbox[{\capbeside\thisfloatsetup{capbesideposition={right,center},capbesidewidth=9cm}}]{figure}[\FBwidth]
 {\hspace{-0.2in} \caption{\footnotesize  {\sc Predicted cumulative number of UV-bright QSO--FRB pairs as a
function of projected angular separation $\theta$ over a common survey
footprint of $\sim$20{,}000\,deg$^2$:}
Both curves assume a purely random (Poisson) spatial distribution,
$N(<\theta) = N_\mathrm{FRB}\,\Sigma_\mathrm{QSO}\,\pi\theta^2$. 
The blue curve shows the current epoch (2026) with
$N_\mathrm{FRB} \approx 100$ arcsecond-localized FRBs at $z < 1$.
The orange curve shows the 2035 forecast with
$N_\mathrm{FRB} = 100{,}000$.
Shaded bands span the combined range of UV-bright QSO surface
density ($\Sigma_\mathrm{QSO} = 0.38$--$0.54\,\mathrm{deg}^{-2}$;
\citealt{Flesch2023}) and the $N_\mathrm{FRB}$ forecast range
($48{,}000$--$135{,}000$; e.g. Sharma et al. 2026).
The horizontal dotted line marks $N = 1$.
The curves assume a static UV-bright QSO catalog and that localized FRBs are distributed uniformly over the survey footprint; see Table~ 1 for the corresponding proper transverse scales
and pair counts at key angular separations.
 \label{pairsepfig}}} 
{\includegraphics[width=9.5cm]{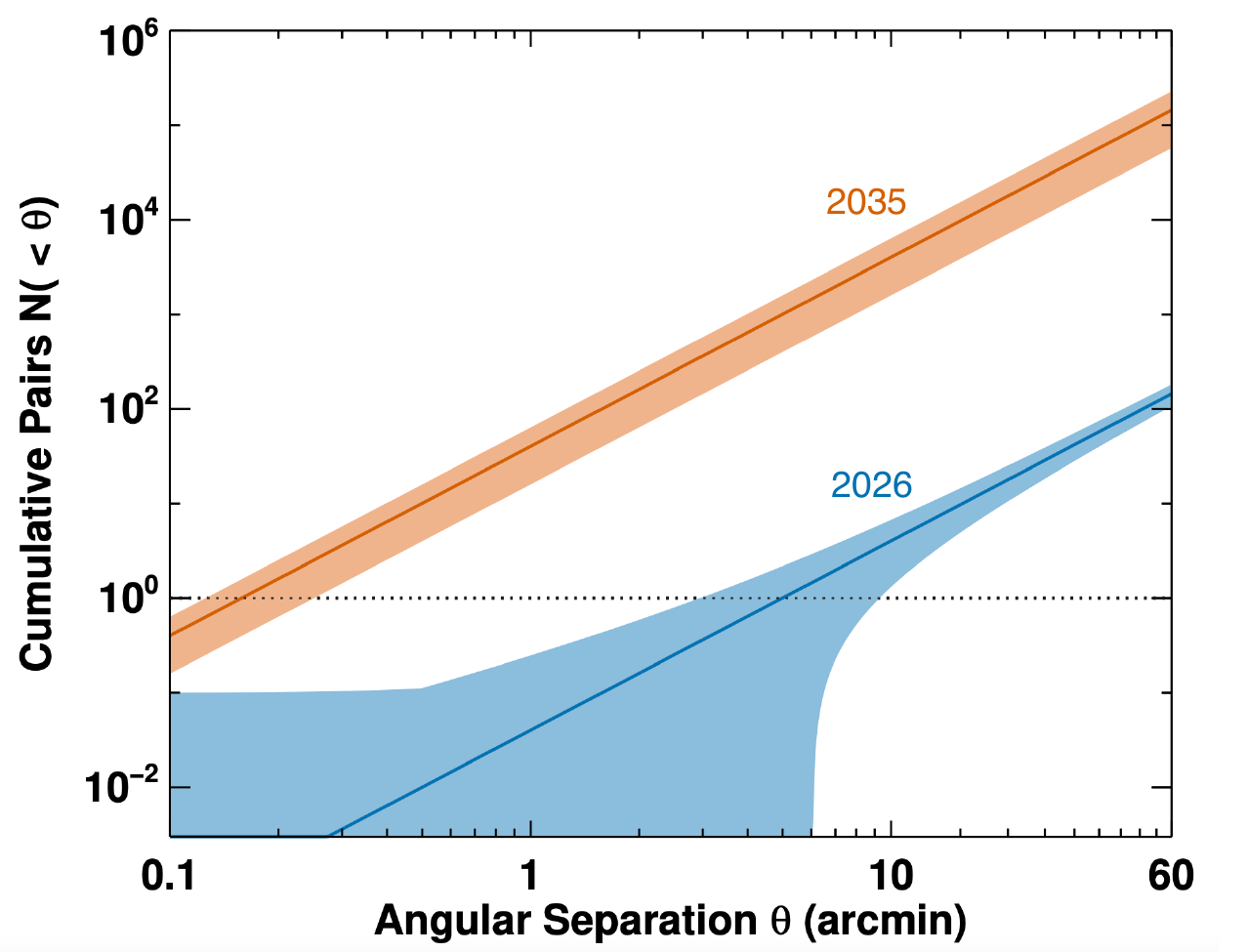}}
\end{centering}
  \vspace{-0.5cm}
\end{figure}
With these inputs, we find fewer than one QSO-FRB pair at $\theta < 1'$ in the present-day sample, borne out by our searches for them in current catalogs. By 2035, this conservative estimate rises to $\sim16-64$ pairs 
at $\theta < 1'$ and thousands of pairs at $\theta < 10'$ (see 
Figure~\ref{pairsepfig} and Table 1). A subset of these pairs may have closely matched redshifts.  

 
\begin{table}[htbp]
\centering
\begin{threeparttable}
\begin{tabular}{lccccc}
\toprule
\multicolumn{6}{c}{\textbf{Projected Physical Scales and Expected UV-bright QSO--FRB Pair Counts}} \\
\midrule
 & \multicolumn{3}{c}{Physical Scale (kpc)} & \multicolumn{2}{c}{$N(<\theta)$ \tnote{a}} \\
\cmidrule(lr){2-4} \cmidrule(lr){5-6}
$\theta$ & $z=0.1$ & $z=0.5$ & $z=1.0$ & \hspace{0.5in} 2026 \tnote{b} & 2035 \tnote{c} \\
\midrule
$20''$  &   37  &  123 &  163 & \hspace{0.5in}$<0.01$ & $2$--$7$   \\
$1'$    &  111  &  370 &  489 & \hspace{0.5in}$<0.05$ & $16$--$64$ \\
$10'$   & 1109  & 3702 & 4894 & \hspace{0.5in} $3$--$5$ & $1600$--$6400$ \\
\bottomrule
\end{tabular}
\begin{tablenotes}
\footnotesize
\item[a] Cumulative pair counts $N(<\theta)$ assuming a Poisson spatial distribution
\item[b] Today ($N_\mathrm{FRB} = 100$, $\Sigma_\mathrm{QSO} = 0.46\,\mathrm{deg}^{-2}$).
\item[c] Range reflects combined uncertainty in $N_\mathrm{FRB}$
  ($48{,}000$--$135{,}000$)
  and $\Sigma_\mathrm{QSO}$ ($0.38$--$0.54\,\mathrm{deg}^{-2}$) for QSOs with FUV $< 19$.
  \vspace{-0.2in}
\end{tablenotes}
\end{threeparttable}
\label{tab:pairs}
\end{table}


\section*{4.\ Science Enabled by Statistical FRB--QSO Pair Surveys}
We outline below how closely paired FRB--QSO sightlines 
will, for the first time, replace the uncertain assumptions 
underlying CGM and IGM baryon budgets with direct, 
phase-resolved measurements. Each science case 
requires a moderate HST/COS exposure of $\lesssim$15 
orbits per target achieving $\mathrm{S/N} \gtrsim 10$ 
per resolution element. The G130M and G160M gratings 
cover $\sim$1150--1775\,\AA\ at $R = 13{,}000$--$17{,}000$ 
($\sim$18--23\,km\,s$^{-1}$ FWHM), resolving individual 
CGM and IGM absorbers to $z \lesssim 1$ in H\,{\sc i} Ly$\alpha$ (and higher order lines, depending on redshift), and metal lines like \ion{C}{II}, \ion{C}{III}, \ion{C}{IV}, \ion{Si}{II}, \ion{Si}{III}, \ion{Si}{IV}, and \ion{O}{VI}. Fewer than $\sim$1,000 of the $\sim$8,000--11,000 UV-bright QSOs accessible over a 20,000\,deg$^2$ footprint \citep{Flesch2023, Monroe2016} have existing HST/COS G130M or G160M spectra with sufficient signal-to-noise for absorption-line science, meaning that the vast majority of future FRB--QSO pairs will require new HST/COS observations to achieve the below goals.  We also note that this program is intrinsically a $z \lesssim 1$, UV endeavor: at higher redshifts localized FRBs will remain too rare for statistical pair studies with deep, high-resolution optical spectra of z$\sim2 -3$ QSOs. 

\subsection*{4.1\ Ionization, Pressure, and Magnetic Fields in the CGM}

The angular separations between FRB and QSO sightlines translate to a wide 
range of physical scales depending on the redshift of any intervening 
structure (Table~1). The most powerful configurations are those in which both 
sightlines pass within $\lesssim 100$\,kpc of each other in the same foreground halo, favoring pairs with $\theta \lesssim 1'$ that intersect the CGM of low-redshift ($z \lesssim 0.3$) galaxies. Because the covering fraction of intervening halos along random sightlines is substantial, this type of configuration will not be rare. The CGM traced by neutral hydrogen extends to $\sim 2\,R_{\rm vir}$ across a wide range of galaxy masses \citep{Wilde2021, Wilde2023}, and by 2035, wide-field surveys such as DESI and its successors will provide precise spectroscopic galaxy redshifts for the field that are required to enable the scientific interpretation. 

FRB dispersion provides a strict upper limit on the free-electron column through 
any intervening halo.  When combined with IGM foreground modeling \citep{Simha2023} and photoionization analysis of the QSO absorption lines \citep[e.g.][]{Werk2014, 
FaermanWerk2023, Qu2018, Zheng2024}, the two sightlines jointly constrain the 
ionization fraction $x_e = n_e/n_{\rm H}$, replacing a dominant systematic in 
CGM mass estimates with a measured boundary condition and enabling a census of how cold cloud and hot halo gas column densities vary as a function of galaxy 
mass and star formation rate. The FRB rotation measure extends this science case further: $\langle B_\parallel \rangle = 
{\rm RM}/{\rm DM}$ yields a density-weighted line-of-sight magnetic field through 
the same plasma, while the pulse broadening timescale $\tau_{\rm scatt}$ 
constrains small-scale electron density fluctuations \citep{McQuinn2014}, a 
potential probe of the non-thermal pressure support that recent theoretical work \citep{Lochhaas2023} 
shows is dynamically significant in $L^*$ halos. The CGM is likely not in simple thermal 
pressure equilibrium \citep{Werk2014}, and cosmic-ray (CR) pressure  may be a significant non-thermal pressure component in $L^*$ halos 
\citep{Butsky2020, Butsky2022}. This is particularly relevant at group-mass scales 
($M \sim 10^{13}$--$10^{14}$\,M$_\odot$), where CR feedback from the central black hole may drive the plasma $\beta$ toward unity at the virial radius \citep{Quataert2025}, a regime in which magnetic fields are dynamically important\footnote{When $\beta \gg 1$, 
thermal pressure dominates and magnetic fields are dynamically negligible; when $\beta \sim 1$, magnetic and thermal pressures are comparable and magnetic 
fields can influence gas structure, suppress instabilities, and guide cosmic-ray 
transport.}. Recent POSSUM observations find magnetic field strengths of several $\mu$G in the intragroup medium of nearby galaxy groups \citep{Anderson2024}, consistent with this prediction.  Measuring $\langle B_\parallel \rangle$ as a function of impact parameter and halo mass across a large FRB--QSO sample will therefore provide the first direct test of whether cosmic-ray pressure plays a role in setting the CGM density and temperature structure across the full range of galaxy and group scales.

\subsection*{4.2\ Apportioning the Cosmic DM Budget}
~The current best constraints on the IGM baryon fraction come from two approaches: fitting the 
statistical DM distribution of $\sim$90 localized FRBs yields $f_{\rm IGM} = 0.76^{+0.10}_{-0.11}$ 
\citep{Connor2025}, while foreground reconstruction 
of the large-scale density field along 8 FRB sightlines yields $f_{\rm IGM} = 0.59^{+0.11}_{-0.10}$ \citep{Khrykin2024}. Both results are effectively global averages. The variance of DM along individual sightlines is dominated by the stochastic intersection of cosmic web filaments, with each sightline traversing $\mathcal{O}(10^2)$ such structures, making it impossible  to assign DM contributions to specific large-scale structures without additional information.  A UV-bright QSO sightline provides exactly that information. The Ly$\alpha$ forest encodes the neutral hydrogen column density as a function of redshift along the same line of sight, identifying gas overdensities at their precise redshifts. Combined with 
the FRB DM, this measurement allows the ionized gas 
contribution of each identified structure to be isolated, 
moving the baryon partition from a global statistical 
inference to a sightline-by-sightline measurement.

The physical scales in Table 1 show that pairs 
at $\theta < 10'$ probe transverse separations of $1.1$--$4.9$\,Mpc 
($z = 0.1$--$1.0$), roughly the characteristic widths of 
cosmic web filaments \citep{AragonCalvo2010a, Cautun2014}. With $1600$--$6400$ such pairs expected by 2035, and each QSO spectrum delivering Ly$\alpha$  and \ion{O}{VI} redshifts, this sample will build a 
three-dimensional, phase-resolved map of the ionized cosmic 
web. The RM of each FRB adds a further constraint at the 
statistical level: while the expected IGM Faraday signal per 
sightline is small ($|\mathrm{RM}| \lesssim$ a few 
rad\,m$^{-2}$), stacking RM measurements across 
$\sim$a hundred of pairs at $\theta < 1'$ with identified 
foreground filaments will place the first observational 
constraints on the mean magnetic field permeating the 
cosmic web as a function of overdensity and redshift.

\subsection*{4.3\ The Milky Way and M31}
The Milky Way CGM is multiphase: a hot, volume-filling 
medium at $T \gtrsim 10^6$\,K, constrained by X-ray observations but with uncertain total electron column, likely coexists with cooler, discrete clouds at $T \sim 10^4$\,K whose large scatter in metallicity  points to multiple physical origins \citep{Faerman2021, Choi2024}. The hot phase likely dominates $\mathrm{DM}_{\rm MW,halo}$, and current constraints treat it as a single sky-averaged upper limit of $52$--$111$\,pc\,cm$^{-3}$ \citep{Cook2023} that folds both contributions together. FRB--QSO pairs can begin to separate them: the UV absorption spectrum traces the cool, photoionized phase cloud by cloud along a specific sightline, while the FRB DM integrates over all ionized gas including the hot, extended component invisible to COS. The difference between the DM-implied electron column and the UV-derived cool-phase contribution is a constraint on the hot phase electron column along that direction, something neither X-ray emission nor UV spectroscopy achieves independently. At $z \sim 0$, 
$\theta < 20''$ probes CGM structures of only a few kpc 
while $\theta \sim 1'$ constrains larger clouds of $\sim$10--20 \, kpc in extent. Assembling such measurements across many sightlines at 
different Galactic latitudes will map the clumpiness, 
phase structure, and hot-to-cool ratio of the MW CGM 
in three dimensions for the first time. For M31, whose 
virial radius subtends $\sim$23$^\circ$, nearly every 
background FRB is a potential halo sightline, and by 2035 
the expected $\sim$100,000 localized FRBs will comfortably 
exceed the $\sim$20,000 needed to constrain M31's radial 
CGM density profile \citep{Kahinga2025}. Hundreds of FRB--QSO pairs within that sample will provide phase-resolved 
constraints on the same profile, supplying the ionization 
fractions and metallicities needed to convert electron 
columns into baryonic masses without the metallicity 
assumptions that currently limit UV-only estimates 
\citep{Lehner2025}.
\subsection*{4.4 Why Not Just Combine Independent Datasets?}
One might reasonably ask whether the science goals described here 
could be achieved by combining the large, independent datasets of
QSO absorption spectra and FRB measurements, without requiring 
paired sightlines. Statistical cross-correlations 
of FRBs with foreground galaxy maps can constrain mean 
DM contributions per halo mass bin \citep[e.g.,][]{Khrykin2024}, 
and stacking analyses of QSO spectra yield mean covering 
fractions and column densities \citep[e.g.,][]{Menard2010}. 
But these approaches average over the physical diversity 
that drives the science cases above. The ionization fraction 
$x_e = n_e/n_{\rm H}$, the CGM pressure balance, and the 
magnetic field strength are likely per-sightline quantities 
that depend on local gas conditions. Their sightline-to-sightline 
variance is not just noise to be beaten down with more data, but rather a signal that provides relevant physical constraints. Coincident sightlines are the only way to 
associate a specific DM and RM with the phase-resolved absorption-line structure of the same gas. This is the same lesson learned from the transition in CGM science from stacking analyses to targeted QSO--galaxy 
pair surveys \citep{Tumlinson2017}: the mean tells you something exists; the individual measurement tells you what it is.
\section*{5.\ The Path to HWO}
Characterizing the diffuse universe, including mapping the CGM and IGM across cosmic time, is one of the science drivers for HWO. The 2030s should not be a scientific dead zone between two great observatories, where the momentum of a 
generation of discoveries with COS is lost. By 2035, arcsecond-localized FRBs will be numerous enough to enable the statistical FRB--QSO pair studies described here for the first time, and HST/COS is the only existing facility before HWO that provides the UV wavelength coverage, spectral resolution, and sensitivity required to carry them out. If {\emph{Hubble}} is boosted, the 2030s can become the 
key decade for high-resolution UV spectroscopy of the 
diffuse universe. FRB--QSO pair surveys like those 
described here, alongside the (hopefully) many other compelling programs proposed in response to this call, will reveal the phases, masses,  clumpiness, and magnetic 
structure of cosmic ecosystems in ways that will define the questions that HWO will be built to answer.

\pagebreak
\setcounter{page}{1}
\pagenumbering{roman} 
\bibliography{frb_qso_whitepaper}
\end{document}